\magnification = \magstep1
\tolerance=99999
\hsize=160true mm
\vsize=247true mm
\voffset=-0.1cm
\hoffset=0.4cm

\def\gee{ \, \lower 1mm\hbox{$\,{\buildrel > \over{\scriptstyle\scriptstyle\sim} }\displaystyle \,$}}
\def\lee{ \, \lower 1mm\hbox{$\,{\buildrel < \over{\scriptstyle\scriptstyle\sim} }\displaystyle \,$}}
\def\|{\partial}
\def\o {\over}
\def\Oo {\displaystyle}
\def\varkappa {{\scriptstyle\partial}\! e}

\nopagenumbers 
\headline={ \ifodd\pageno\rightheadline \else\rightheadline\fi }
\def\rightheadline{\it\hfil        { }  \hfil\folio }

\newcount\notenumber

\def\note{\advance\notenumber by 1
\footnote{$^{\the\notenumber )}$}}

% Font defs
\font\hdf=cmr8 scaled \magstep0
%\font\figf=cmr9 scaled \magstep0
%\font\big=cmr14 scaled \magstep0
%\font\epf=cmti10 scaled \magstep0
%\rm

\centerline{\bf DISSIPATION INSTABILITIES IN THE ACCRETION DISK }

\bigskip

\centerline{ A.V. KHOPERSKOV and S.S. KHRAPOV}
\centerline{\it Department of Theoretical Physics, Volgograde State University,  
Volgograd, Russia}

\bigskip
\noindent
{\hdf {\bf Abstract.}  
The model of a geometrically thin gaseous disk in the external 
gravitational potential is considered. 
The dinamics of small nonaxisymmetric perturbations in 
the plane of the accretion disk with dissipative effects is investigated.
It is showed, that conditions of development and 
parameters of unstable oscillation modes in the opticaly thick accretion disk 
are strongly depended on the models of viscosity and opacity.
}

\bigskip

The possibility of the development of various types of instabilities in 
a thin gaseous disks is very attractive one for understanding different aspects 
of the accretion-disk (AD) phenomenon. 
In order to explain the required large  values of dissipative coefficients, 
the concept of the turbulent viscosity has been used, which may be caused 
by the developed turbulence 
of a gaseous medium arising from the loss of stability. 
On the other hand, 
the nonlinear evolution of unstable oscillation modes may be responsible for 
many nonstationary phenomena in accreting systems.

There are four unstable oscillation modes in the framework  
of the standard $\alpha$-model of an accretion disk [1]. 
Two of then are acoustic [6-11], one is viscous and the other one 
is thermal [2-11]. 
A distinctive feature of the dynamic viscosity 
$\eta = \sigma \nu$ in the model of the accretion disks is it's dependence 
on the surfase density $\sigma$ and on the disk half-thicness $h$ 
($\nu$ is the kinematic viscosity). 
The perturbation of the dynamic viscosity $\tilde \eta$ is 
responsible for the formation of all unstable four modes.
However, in all the above-cited papers on the dynamic of linear perturbation 
is suggested that viscosity simultaneously changes with changing of   
accretion-disk parameters.

In our present work, we extend our research 
to time delay of the viscosity influence on the 
thermal, viscous and acoustic instabilities.
In the construction of a different viscous models AD 
is suggested that viscosity caused by the developed turbulence 
$\eta \sim \sigma u_t \ell_t$, 
where $u_t$ and $\ell_t$ is the characteristic velocity of the 
large-scale turbulent pulsations and it's characteristic size, 
respectively [12].
The fundamental energy is contained in the large-scale pulsations, however 
dissipation of the energy is the case in the small-scale pulsations. 
By this means, as the local conditions changes in the accretion disk, 
there are two factors, 
which involves the change delay of the turbulent viscosity. 
Firstly, since the turbulence may be caused by the nonlinear evolution 
of unstable oscillation modes, that for formation 
of the developed turbulence is required 
of the characteristic time $\tau_1$. 
The first approximation may be thought of as $\tau_1$ is proportional to 
the build-up time of instabilities. 
Secondly, there are delay $\tau_2$, associated with 
transfer of an energy from large-scale pulsations to small-scale pulsations. 
Hence, as accretion-disk parametrs changes 
(of the temperature and density), value of the actual viscosity 
time delay from instant value of the dynamic viscosity $\eta_\ast$ 
by characteristic time $\tau = \tau_1 + \tau_2$. 
For standard $\alpha$-model AD $\eta_\ast \sim \alpha \sigma \Omega h^2$,  
where $\Omega$ is the Keplerian angular velocity. 
The first approximation law of relaxation viscosity $\eta$ 
to value $\eta_\ast$ may be writen in the following form:
$$
\Oo{d\eta \o dt} = {\eta_\ast - \eta \o \tau}\,, \eqno(1)
$$ 

We restrict ourselves to the case of small perturbation with $kr \gg 1$ 
and $m/r \ll k$ (k is the radial wavenumber and $m$ is azimuthal wavenumber), 
which allows us to use WKB approximation and seek the solution in the form 
$$
\tilde f = f_1 exp(-i\omega t + ikr + im\varphi)\,\, , \eqno(2)
$$
where $\omega$ is the complex frequency of a mode. Equilibrium 
quantities are denoted by the subscript "0".

\bigskip
\bigskip
\centerline{\bf The influence of the delay of the viscosity.}
\medskip

In the general case $\tau > 0$ we obtain the five-order dispersion relation.
This is equation describes five oscillatory modes. 
Four of them were considered previously at $\tau = 0$ [11]. 
The inclusion of the delay $\tau >0$ gives rise to new oscillatory mode 
is the second-viscous mode, 
beside  ${\rm Re}(\omega) = 0$ and ${\rm Im}(\omega) < 0$ 
at any values of another parameters. 
Rewriting equation (1) in terms of the result (2) we obtain 
in the linear approximation :
$$
{\eta_1 \o \eta_0} = {1\o 1-i\omega \tau}\,{\eta_{\ast 1} \o \eta_0}\,\, , 
\eqno(3)
$$
where \ \ 
$\Oo{
{\eta_{\ast 1} \o \eta_0} = {\sigma_1 \o \sigma_0} + {\nu_{\ast 1}\o \nu_0} = 
(1+\delta_\sigma )\,{\sigma_1 \o \sigma_0} + \delta_h {h_1\o h_0}\,,\,\,\, 
\delta_\sigma=\left({d\ln \nu_\ast \o d \ln\sigma}\right)_0\,,\,\,\,  
\delta_h=\left({d\ln \nu_\ast \o d \ln h}\right)_0  }$  [11].

For a standard $\alpha$-model accretion disk in the 
radiation-pressure-dominated region the increment of all four unstable 
modes decreases, with the increase of characteristic time of the delay $\tau$, 
as indicated by fig.~1. 
Stabilization of the thermal and viscous 
oscillation modes (${\rm Im}(\omega) \lee 0$)
occurs at $\tau \gee 100/\Omega$, while the acoustic mode tend to become 
stable at $\tau \simeq 1/\Omega$.

\bigskip
\centerline{\bf The influence of the opacity }
\medskip

The conditions of development for unstable oscillation modes very strongly 
depend on the model of opacity.  
We assume that opacity  $\bar \kappa$ is a function of $\sigma$ and $h$ 
(in other words, of density and temperature). The linear approximation yelds
$$
{\bar \kappa_1 \o \bar \kappa_0} = \Delta_\sigma {\sigma_1 \o \sigma_0} + 
\Delta_h {h_1\o h_0}\, ,
\quad\ \left\{\Delta_\sigma=\left({d\ln \bar \kappa \o d \ln\sigma}\right)_0\,, 
\ \ \Delta_h=\left({d\ln \bar \kappa \o d \ln h}\right)_0  \right\}\,. 
\eqno (4)
$$

In the case of the Thomson scattering  
$\bar \kappa = \bar \kappa_{es} = 0,4 {\rm sm^2/g}$ 
($\Delta_\sigma = 0$, $\Delta_h = 0$), 
and for Kramers law $\bar \kappa = \bar \kappa_{ff} \propto \rho T^{-7/2}$ 
($\Delta_\sigma = 1$, 
$\Delta_h = -8$). 
For low temperature protoplanetary disks  
$\bar \kappa \propto  T^2$ [13] 
($\Delta_\sigma = 0$, 
$\Delta_h = 4$). 
In the "hot" limit of Foulkner model [14] 
$\bar \kappa \propto \rho T^{-5/2}$ 
($\Oo{\Delta_\sigma = {2+\beta_0\o 2(1+3\beta_0)}}$, 
$\Oo{\Delta_h = -{12+\beta_0\o 2(1+3\beta_0)}}$), 
but in the "cold" limit $\bar \kappa \propto \rho^{1/3} T^{10}$ 
($\Delta_\sigma = 1/3$, 
$\Delta_h = 59/3$). Here $\beta_0 = P_{0\,rad}/(P_{0\,rad} + P_{0\,gas})$, 
$P_{gas}$ is the gas pressure, $P_{rad}$ is the radiation pressure.

As indicated by fig.~2 the thermal mode of the oscillation become unstable 
even in the case $\beta_0 = 0$ at  $\Delta_\sigma < 0$ and $\Delta_h > 0$. 
Stabilization of the acoustic-mode occurs at a high negative value 
$\Delta_\sigma$.  
However, in the framework of standard model AD [1] acoustic modes 
of the oscillations prove to be unstable both in the inner 
radiation-dominated region ($P_{rad} \gg P_{gas}$, 
$\bar \kappa = \bar \kappa_{es}$), 
and in the external gaseous region ($P_{gas} \gg P_{rad}$, 
$\bar \kappa = \bar \kappa_{ff}$).

\bigskip
\centerline{\bf The influence of the viscosity}
\medskip

As the second (elastic) viscosity $\mu_0$ increases, the increment 
of the sound waves 
decreases until the imaginary part vanishes at $\mu_0 = \mu_{0\, crit}$. 
The perturbation decay (${\rm Im}(\omega) < 0$) when $\mu_0 > \mu_{0\, crit}$. 
As shown in the work [11] the quantity $\mu_{0\, crit}$ 
only weakly  depends on $\beta_0$, so that $\mu_{0\, crit}= 2 \div 3 \nu_0$.

The increments and conditions for instability development very strongly 
depend on the parameters $\delta_\sigma$ and $\delta_h$. 
By now there are many models of turbulent viscosity 
with differents values $\delta_\sigma$, $\delta_h$.
For standard $\alpha$-model $W_{r\varphi}=-\alpha p$ [1]  and 
$\delta_\sigma =0$, $\delta_h=2$.
For modifications of $\alpha$-models $W_{r\varphi}=-\alpha (p_g/p)^{N /2}p$ 
($N={\rm const}$) [2,4,5] and   
$\Oo{\delta_\sigma = {N\beta_0 \o 2(1+3\beta_0)}} $; 
$\Oo{\delta_h = {4+\beta_0(12-7N) \o 2(1+3\beta_0)}}$. 

The functions ${\rm Im}[\omega(\delta_\sigma)]$ and ${\rm Im}[\omega(\delta_h)]$ 
are in qualitative agreement  with
${\rm Im}[\omega(\Delta_\sigma)]$ and ${\rm Im}[\omega(\Delta_h)]$ accordingly.
In the terms of the mentioned models of turbulent viscosity, both 
acoustic modes of the oscillations are unstable 
at $\mu_0 < \mu_{0\,crit}$, 
because for all these models $\delta_\sigma > 0$ and $\delta_h > 0$. 
The acoustic oscillations stabilize at large negative values of $\delta_\sigma$. 

\bigskip
\centerline{\bf Conclusion }
\medskip

A linear analysis has been performed to examine the radial-azimuthal 
instability of accretion disks. The main results are as follows:

First, the existence of unstable modes is associated completely with  
the perturbation  
of dynamic viscosity $\eta=\sigma\nu$ and, consequently, it is 
determined by the dependence of $\nu(\sigma,h)$. 
If we set $\tilde \eta \equiv 0$, 
all four oscillatory modes (including the thermal and viscous modes in the 
radiation-dominated limit) would decay with decrement 
${\rm Im}(\hat \omega) \sim -\nu_0 k^2$. 

Second, the azimuthal perturbation wavenumber $m$ 
in terms of the WKB approximation for a short-wave perturbations with 
$kr \gg 1$ not affected by the increments of unstable modes and 
that is responsible for Dopplershift of the frequency 
$\hat \omega = \omega - m\,\Omega$. 

Third, in the radiation-pressure-dominated accretion disk 
the thermal and viscous unstable modes stabilize at 
large values of the characteristic time scale of the delay $\tau$, 
while the acoustic mode tend to become stable at small values $\tau$.

Finally, the conditions of development and 
parameters of the thermal, viscous and acoustic instabilities 
in the geometrically thin and opticaly thick accretion disk are strongly
depended on the models of viscosity and opacity. 
In the gas-pressure-dominated accretion disk the thermal mode tend to 
become unstable with decreases of the values $\delta_\sigma \,,\Delta_\sigma$ 
and with increases of the values $\delta_h \,, \Delta_h$.
The acoustic mode at any gas-to-radiation pressure ratio tend to become 
stable with decreases of the values $\delta_\sigma \,,\Delta_\sigma$. 
An additional point to emphasize is that 
the sound waves stabilize at large values of the second (elastic) viscosity.

\bigskip
\centerline{\bf Acknowledgments}
\medskip

We wish to thank V.V. Mustsevoi, V.V. Levi and I.G. Kovalenko for 
discussion of the results and the reviewers for useful remarks.

\vfill\eject

\centerline{\bf References}
\medskip

%{\hdf
\noindent
1. Shakura N.I. $\&$ Sunyaev R.A., 1973, A$\&$A 24, 337

\noindent
2. Lightman A.P. $\&$ Eardley D.M., 1974, ApJ 187, L1 

\noindent
3. Shakura N.I. $\&$ Sunyaev R.A., 1976, MNRAS 175, 613

\noindent
4. Taam R.E. $\&$ Lin D.N.C., 1984, ApJ 287, 761

\noindent
5. Szuszkiewicz E., 1990, MNRAS 244, 377

\noindent
6. Wallinder F.H., 1990, A$\&$A 237, 270

\noindent
7. Wallinder F.H., 1991, MNRAS 253, 184

\noindent
8. Wallinder F.H., 1991, A$\&$A 249, 107

\noindent
9. Okuda T. $\&$ Mineshige S., 1991, MNRAS 249, 684

\noindent
10. Wu X-B. $\&$ Yang L-T., 1994, ApJ 432, 672

\noindent
11. Khoperskov A.V. $\&$ Khrapov S.S., 1995, Astronomy Letters 21, N3, 347

\noindent
12. Landau L.D. $\&$ Lifshits E.M., 1986, Gidrodynamic, W: Science 

\noindent
13. Lin D.N.C., 1981, ApJ 246, 972 

\noindent
14. Foulkner J., Lin D.N.C. $\&$ Papaloirou J., 1983, MNRAS 205, 359

%}

\vfill\eject

\centerline{\bf Figure Captions}
\bigskip

\noindent
Fig.~1. Dependence of the imaginary part of the frequency in terms 
of the angular velocity ${\rm Im}(\omega )/\Omega$ 
on the without dimensional wavenumber $kh_0$ a),b) 
and on the without dimensional time of delay $\tau \Omega$ c),d). 

\bigskip
\noindent
Fig.~2. Dependence of the imaginary part of the frequency in terms of 
the angular velocity ${\rm Im}(\omega )/\Omega$ 
of acoustic, thermal and viscous oscillation modes 
on the value parameter $\Delta_\sigma$ a),b) and 
on the value parameter $\Delta_h$ c),d) at $kh_0 = 1$. 
The solid and short dashed lines represent $\beta_0 = 0$ and 
$\beta_0 = 1$, respectively.

\end